\documentclass[preprintnumbers,showkeys,showpacs,byrevtex]{revtex4}
\usepackage{amsmath,amsfonts,amssymb,amscd,amsxtra,amsthm}
\usepackage{graphicx}
\usepackage{multirow}
\usepackage{bm}
\usepackage{epstopdf}
\begin{document}
\preprint{KIAS-P12029}
\title{Parton-distribution functions for the pion and kaon in the gauge-invariant\\
nonlocal chiral-quark model}
\author{Seung-il Nam}
\email[E-mail: ]{sinam@kias.re.kr}
\affiliation{School of Physics, Korea Institute for Advanced Study (KIAS), Seoul 130-722, Republic of Korea}
\date{\today}
\begin{abstract}
We investigate the parton-distribution functions (PDFs) for the positively charged pion and kaon at a low-renormalization scale $\sim1$ GeV. To this end, we employ the gauge-invariant effective chiral action from the nonlocal chiral-quark model, resulting in that the vector currents are conserved.  All the model parameters are determined phenomenologically with the normalization condition for PDF and the empirical values for the pseudoscalar-meson weak-decay constants. We consider the momentum dependence of the effective quark mass properly within the model calculations. It turns out that the leading local contribution provides about $70\%$ of the total strength for PDF, whereas the nonlocal one that is newly taken into account in this work for the gauge invariance does the rest. High-$Q^2$ evolution to $27\,\mathrm{GeV}^2$ is performed for valance-quark distribution function (VQDF), using the DGALP equation. The  moments for the pion and kaon VQDFs are also computed. The numerical results are compared with the empirical data and theoretical estimations, and show qualitatively agreement with them. 
\end{abstract}
\maketitle
\section{Introduction}
It has been well-known that the high-energy scattering processes, such as the inclusive or exclusive production ones, are very useful tools to study Quantum ChromoDynamics (QCD)~\cite{Collins:1992kk,Mulders:1995dh,Boer:1997nt,Anselmino:1994tv,Anselmino:2008jk,Christova:2006qs,Anselmino:2007fs,Bacchetta:2006tn,Efremov:2006qm,Collins:2005ie,Ji:2004wu}. In such processes, the Drell-Yan (DY) one for instance, the scattering amplitudes consist of short and long range QCD interactions simultaneously. By the virtue of the factorization theorem, the first can be studied via perturbative QCD (pQCD), whereas one is able to investigate the second, that signals the nontrivial structures of the hadrons involved, by various nonperturbative approaches. Note that those nonperturbative quantities can be defined by the parton-distribution amplitude (PDA), parton-distribution function (PDF), fragmentation function (FF), generalized parton-distribution function (GPD), and so on, depending on different scattering processes.  Among them, PDFs for the pseudoscalar (PS) mesons, $\phi=(\pi,\,K)$, are of importance to understand the nonperturbative structure of the mesons, which play a crucial role in the low-energy QCD. In Ref.~\cite{Shigetani:1993dx}, employing the Nambu--Jona-Lasinio (NJL) model, PDF and valance-quark distribution function (VQDF) for the pion and kaon were derived from the forward scattering amplitude of a virtual photon from a pion target in the Bjorken limit. The obtained results at a low renormalization scale $\sim1$ GeV were evolved to a high $Q^2$ by using the Altarelli-Parisi equation. Similarly, in Ref.~\cite{Davidson:1994uv}, the SU(3) NJL model was applied for PDF for the pion, kaon and eta mesons with scalar and pseudoscalar couplings. They obtained, in the chiral limit, the simple result $f_\phi(x) = f_\phi(x) = \theta(x)\theta(1 -x)$ for the structure functions, satisfying the gauge invariance for the vector current. In Ref.~\cite{Weigel:1999pc}, the chiral quark model (ChQM) was used for the pion PDF, resulting in that the Pauli-Villars regularization scheme is most suitable for both the anomaly structure of QCD and the leading scaling behavior of PDF in the Bjorken limit. The pion PDF was computed using the instanton-liquid model, suggesting an analytic expression for a general vertex function and satisfying, in a gauge invariant approach ~\cite{Dorokhov:2000gu}.
The NJL-jet model was also employed to compute PDF as well as FF by the cut diagrams in Ref.~\cite{Ito:2009zc}. A parameter-free prediction for the ratio $u_K(x)/u_\pi(x)$ was given in Ref.~\cite{Nguyen:2011jy} using the rainbow-ladder truncation for the Dyson-Schwinger (DS) equations, reproducing the DY data.  Extraction of the pion PDF was performed in a next-to-leading
order analysis from Fermilab E-615 pionic DY data, observing that the high-$x$ dependence is different from that of the leading order analysis, whereas it does not match with the perturbative QCD (pQCD) and DS calculations~\cite{Wijesooriya:2005ir}. The lowest three non-trivial moments, calculated by the lattice-QCD (LQCD) simulation, were in good agreement with existing data at the physical pion mass~\cite{Detmold:2003tm}. In Ref.~\cite{Daniel:1990ah}, the two-flavor Wilson-fermion was used for LQCD simulation, observing smallness of relevant moments in comparison with other theoretical models. Using QCD sum rules, the pion PDF was investigated taking into account the nonlocal condensates, paying attention on the bilocal power corrections~\cite{Belitsky:1996vh}. In the previous works~\cite{Nam:2011hg,Nam:2012af}, we also computed unpolarized FF first then converted it into PDF by using the Drell-Levi-Yan (DLY) relation~\cite{Drell:1969jm}, employing the simplified nonlocal chiral-quark model (NLChQM), although this duality or analytic continuation according to the DLY relation is not generally satisfied especially for $T$-odd PDFs. The numerical results turned out to be compatible with available empirical data after the DGALP (Dokshitzer-Gribov-Lipatov-Altarelli-Parisi) evolution for high-$Q^2$ values. 

We note that, although the previous works showed qualitatively reasonable results for PDF and FF~\cite{Nam:2011hg,Nam:2012af}, from a theoretical point of view, there were some issues to be improved: 1) The vector-current conservation (or gauge invariance) of the matrix element for PDF was not taken into account, i.e. only the simple {\it local} contribution was computed. 2) Moreover, the numerical calculations were simplified to a certain extent by assuming that the constituent-quark masses inside the relevant matrix element for PDF are constant, not momentum dependent, except for the quark-PS-meson couplings, although this simplification helps to understand the analytic structure of PDF in the model  and reduces the difficulties much in the numerical calculations.  3) Above all, PDF was converted from FF via the DLY, not a direct calculation. Hence, in the present work, we would like to improve all the issues mentioned above in the same model. Our strategy in the present work is as follows:
\begin{enumerate}
\item PDF is directly computed by the gauge-invariant effective chiral action (EChA) of NLChQM, satisfying the vector-current conservation. This is a similar approach as in Refs.~\cite{Nam:2006sx,Dorokhov:2000gu}.
\item All the momentum dependences in the constituent-quark masses $M_f$, in which the subscript $f$ stands for the quark flavor, are strictly taken into account: $M_f\equiv M_f(k^2)$. One can refer for more details in Refs.~\cite{Praszalowicz:2001wy,Nam:2006au}.
\item Relevant model parameters are determined by satisfying the theoretical condition and experimental information, i.e. the PDF normalization condition, and empirical values for the pion and kaon weak-decay constants. 
\end{enumerate}

As already observed in Refs.~\cite{Nam:2006sx}, it turns out from the numerical results in the present work that the {\it nonlocal} contribution, which are newly considered here and necessary to preserve the gauge invariance,  provides about $30\%$ of the total strength for PDF. Similar tendency was also observed in Refs.~\cite{Dorokhov:1998up,Dorokhov:2000gu} in the single-instanton model. Moreover, those contributions play the role of broadening PDF. Interestingly, careful treatment of the momentum dependences in the effective quark masses does make noticeable changes in the numerical results in comparison to those with the simplification, i.e. $M_f(k)\to\mathrm{constant}$, if we compare them with the previous calculations. Considering the normalization condition for PDF and $F_\pi=93.2$ MeV as an input, we determine one of the model parameters, the constituent-quark mass at zero virtuality, to be $M_0\approx300$ MeV, once we choose the model renormalization scale as $\mu\approx1$ GeV phenomenologically as in Ref.~\cite{Nam:2006sx}. From this parameter set, the kaon weak decay constant is obtained as $F_K\approx121.7$ MeV, which is only about $7\%$ larger than its empirical value, $113.4$ MeV. From the computed PDF for the pion $(\pi^+)$ and kaon $(K^+)$, we parameterize them into a simple analytic form, which was suggested in various works for extracting empirical PDF from the experimental data~\cite{Conway:1989fs,Sutton:1991ay}. Employing the DGLAP equation for high $Q^2$ evolutions for them, we compare our results with empirical data, showing considerable agreement with them. We also present the numerical results for the ratio of $u_K/u_\pi$ and the moments of the pion and kaon VQDFs in comparison with the data~\cite{Sutton:1991ay,Aicher:2010cb,Martinelli:1987bh} to verify the relevance of the present theoretical work. 

The present work is organized as follows: In Section II, we briefly introduce the present theoretical framework, defining PDF and VQDF for the PS mesons. The numerical results and related discussions will be given in Section III. The last Section is devoted to summary, conclusion, and future perspectives. 

\section{Theoretical framework}
\begin{figure}[t]
\includegraphics[width=13cm]{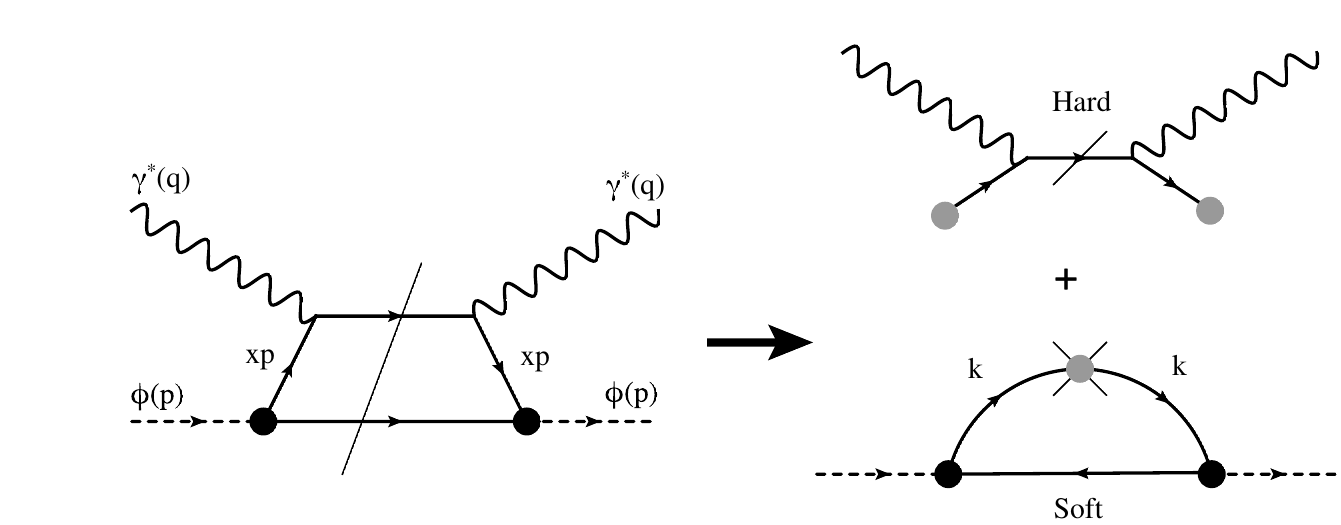}
\caption{The left diagram for the hadronic tensor $W_{\mu\nu}$ in Eq.~(\ref{eq:HADRON}) for the PS meson $\phi$  with its on-shell momentum $p$, satisfying $p^2=m^2_\phi$. The solid, dash, and wavy lines indicate the pseudoscalar (PS) mesons, quarks, and virtual photon, respectively. $x$ stands for the momentum fraction. The thin slashes denote the discontinuity for the imaginary parts. The right diagrams show the factorized ones as the {\it hard} and {\it soft} contributions. The soft part represents PDF with the loop momentum $k$. The black blobs stands for the nonlocal quark-PS-meson vertex, whereas the gray one in the soft diagram for the local operator for PDF in Eq.~(\ref{eq:MO}).}       
\label{FIG1}
\end{figure}

In this Section, we explain briefly the definition for PDF for the PS mesons, i.e. pion and kaon. For definiteness, we choose only positively charged pion and kaon, and assign them as $\pi^+\sim u\bar{d}\equiv \pi$ and $K^+\sim u\bar{s}\equiv K$ hereafter for simplicity. For other isospin states, it is straightforward to compute the multiplicable factor considering the isospin symmetry. In what follows, we want to explain how to define and calculate PDF. In general, information for the parton distribution function can be extracted from the $(\pi\,\mathrm{or}\,K)$-$N$ scattering using the DY process or deeply virtual Compton scattering (DVCS) of the PS mesons. The information of the PS-meson  PDF can be represented by  the hadronic tensor $W_{\mu\nu}$ as depicted in the left of Figure~\ref{FIG1}, which is so-called the handbag diagram in the forward Compton scattering,  and it is defined as a function of the Bjorken $x$ by~\cite{Shigetani:1993dx,Frederico:1994dx} 
\begin{equation}
\label{eq:HADRON}
W_{\mu\nu}=-\left(g_{\mu\nu}-\frac{q_\mu q_\nu}{q^2} \right)F_1(x)+
\frac{1}{m_\phi\nu}\left(p_\mu-\frac{p\cdot q}{q^2} q_\mu\right)
\left(p_\nu-\frac{p\cdot q}{q^2} q_\nu\right)F_2(x),\,\,\,\,x=-\frac{q^2}{2m_\phi\nu},
\end{equation}
where the four momenta $p$ and $q$ stand for those of the PS-meson and virtual photon as shown in the left of Figure~\ref{FIG1}. $m_\phi$ and $\nu$ denote the PS-meson mass and $(p\cdot q)/m_\phi$, respectively. The structure functions in Eq.~(\ref{eq:HADRON}) read
\begin{equation}
\label{eq:F12}
F_1(x)=x\sum_f e^2_f[f_\phi(x)+\bar{f}_\phi(x)],\,\,\,\,
F_2(x)=\frac{F_1(x)}{2x}.
\end{equation}
Here, $f_\phi(x)$ and $\bar{f}_\phi(x)$ indicate PDF for the quark and antiquark, while $e_f$ denotes the electric charge for the quark. Thus, by measuring $F_1(x)$ by experiments, one can extract PDF at a certain $Q^2$ value. There can be two ways to compute PDF based on nonperturbative QCD techniques. By defining the interaction vertices in the handbag diagram, i.e. the $qq\pi$ and $qq\gamma$ vertices, from effective models manifesting the spontaneous breakdown of chiral symmetry (SBCS), such as the NJL model~\cite{Shigetani:1993dx}, light-front formalism~\cite{Frederico:1994dx} and so on, one can compute   directly  the handbag diagram in the left of Figure~\ref{FIG1}, using the optical theorem and the Bjorken limit $Q^2\to\infty$, then determine $F_1(x)$ and PDF via Eqs.~(\ref{eq:HADRON}) and (\ref{eq:F12})~\cite{Shigetani:1993dx,Frederico:1994dx}. It turned out that this method for computing PDF is equivalent to the operator-product-expansion (OPE) method, which will be explained below~\cite{DeRujula:1980qh}.

The other way is to extract the {\it soft} part for PDF from the hadron tensor $W_{\mu\nu}$ in terms of the OPE technique considering the factorization theorem~\cite{Dorokhov:2000gu,Ito:2009zc}. Due to the OPE, the handbag diagram can be separated by the convolution of the hard  and soft parts, and the separated diagrams are depicted in the right of Figure~\ref{FIG1}. The hard one is represented by Wilson coefficients, whereas the soft one by the matrix elements of local operators in a nonperturbative manner. In the present work, we employ this method with help of the gauge-invariant NLChQM. Now, we want to focus on the effective local operators for OPE, categorized by its twist and denoted by the gray blob of the soft diagram in Figure~\ref{FIG1}. Since higher twists are suppressed at high $Q^2$, we will take into account only the twist-$2$ operator for PDF. The relevant matrix element of the local operator sandwiched by the PS-meson states can be related to the $m$-th moments of PDF as follows~\cite{Dorokhov:2000gu}: 
\begin{equation}
\label{eq:MO}
\frac{i^{m+1}}{2}\langle \phi(p)|\bar{q}_f(0)\rlap{/}{n}(n\cdot D)^{m}q_f(0)|\phi(p)\rangle
=\int^1_0dx\, x^{m}f_\phi(x)\equiv\langle x^m\rangle_\phi,
\end{equation}
where $D_\mu$ and $n_\mu$ denote the covariant derivative and light-like vector, which satisfies $n^2=0$, respectively. Note that the soft part is characterized by a renormalization scale $\mu$, at which nonperturbative natures are determined. Hence, we will take $\mu$ implicitly in PDF hereafter. The $0$-th moment with $m=0$ becomes for instance:
\begin{equation}
\label{eq:MO1}
\frac{i}{2}\langle \phi(p)|\bar{q}_f(0)\rlap{/}{n}q_f(0)|\phi(p)\rangle
=\int^1_0dx\, f_\phi(x)=1.
\end{equation}
Note that Eq.~(\ref{eq:MO1}) must be unity to satisfy the normalization condition for VQDF (or PDF). Readers can refer to Eqs.~(\ref{eq:RE}) and (\ref{eq:RE2}) for the relations between various PDFs and VQDFs for $\pi$ and $K$. Alternatively, PDF can be also defined by~\cite{Jaffe:1983hp,Ellis:1991qj,Nguyen:2011jy}:
\begin{equation}
\label{eq:PDFDEFI}
f_\phi(x)=\frac{i}{4\pi}\int\,d\eta\,e^{i(xp)\cdot(\eta n)}
\langle\phi(p)|\bar{q}_f(\eta n)\rlap{/}{n}q_f(0)|\phi(p)\rangle.
\end{equation}
One can easily show that Eq.~(\ref{eq:PDFDEFI}) is equivalent to Eq.~(\ref{eq:MO1}). The displacement was assigned by $\eta$ in Eq.~(\ref{eq:MO1}). It can be understood by comparing the handbag and soft diagrams in Figure~\ref{FIG1}, the loop momentum $k$ must satisfy the following condition, $x\,p\cdot n=k\cdot n$, after the factorization. This condition can be expressed in terms of a delta function in the loop integral over $k$, i.e. $\delta(k\cdot n-x\,p\cdot n)$, which corresponds to the effective composite local operator. At the same time, the momentum fraction $x$ is also defined by $x=(k\cdot n)/(p\cdot n)$. Here, the light-like vector $n$ picks up the momentum (spatial) component of a vector, i.e. $n\cdot v=v^+$ in the light-cone coordinate. In our theoretical framework, the following relations are satisfied for the positively charged pion and kaon:
\begin{equation}
\label{eq:RE}
u_\pi(x)=\bar{d}_\pi(x),\,\,\,\,\bar{u}_\pi(x)=d_\pi(x)=s_\pi(x)=\bar{s}_\pi(x)=0,
\,\,\,\,\bar{u}_K(x)=\bar{d}_K(x)=d_K(x)=s_K(x)=0.
\end{equation}
Here, the first relation in Eq.~(\ref{eq:RE}) comes from the isospin symmetry, $m_u=m_d$, assumed in the present work. According to Eq.~(\ref{eq:RE}), we have the following relations between PDFs and VQDFs:
\begin{equation}
\label{eq:RE2}
u^V_\pi(x)=u_\pi(x),\,\,\,\,
d^V_\pi(x)=-\bar{d}_\pi(x)=-u_\pi(x),\,\,\,\,
s^V_\pi(x)=0,\,\,\,\,
u^V_K(x)=u_K(x),\,\,\,\,
d^V_K(x)=0,\,\,\,\,
s^V_K(x)=-\bar{s}_K(x),\,\,\,\,
\end{equation}

Now, we are in a position to compute the soft diagram representing PDF using an effective model. For that end, we employ NLChQM~\cite{Nam:2006sx,Praszalowicz:2001wy}, being motivated by the instanton physics, which is properly defined in Euclidean space~\cite{Shuryak:1982dp,Diakonov:2002fq}. The dilute instanton-liquid model is characterized by the nontrivial interactions between the quarks and (anti)instantons, which represents the nonperturbative QCD configuration, being characterized by the average inter-(anti)instanton distance $\bar{R}\sim1$ fm and (anti)instanton size $\bar{\rho}\sim1/3$ fm. According to the nontrivial interactions, the quarks acquire their effective masses, resulting in the spontaneous breakdown of chiral symmetry (SBCS). The effective action of the instanton model can be bosonized, resulting in an effective chiral model with the quark and PS-meson degrees of freedom~\cite{Diakonov:2002fq}. This effective chiral model possesses very interesting features: As mentioned, the dynamically-generated effective quark mass becomes $(300\sim400)$ MeV to satisfy phenomenology and becomes a decreasing function of the momentum transfer. Hence, it plays the role of a UV regulator by construction, and the interaction strengths between the quarks and PS mesons becomes nonlocal, i.e. momentum dependent. Since we are interested in the physical quantities in Minkowski space, we perform the Wick rotation for the instanton model, i.e. $t\to i\tau$, then, obtain NLChQM, which resembles the nonlocal NJL model in many aspects. Although there is no firm theoretical proof for the validity of this analytic continuation between the instanton model and NLChQM, from a practical point of view, we employ NLChQM in the present work keeping the typical features of the instanton model. 

Taking into account all the ingredients discussed above, the effective chiral action (EChA) of NLChQM can be written in a neat form:
\begin{equation}
\label{eq:ECA}
\mathcal{S}_\mathrm{eff}[\phi,m_f;\mu]=-i\mathrm{Sp}\ln\left[i\rlap{/}{\partial}
-\hat{m}_f-\sqrt{M_f(i\rlap{/}{\partial})}U_5\sqrt{M_f(i\rlap{/}{\partial})} \right],
\end{equation}
where $\mathrm{Sp}$ stands for the functional trace, $\mathrm{Tr}_{c,f,\gamma}\langle x|\cdots|x\rangle$, in which $(c,f,\gamma)$ denote the (color, flavor, Lorentz) indices, respectively, whereas $\hat{m}_f$ indicates the current-quark mass matrix, $\mathrm{diag}(m_u,m_d,m_s)$. Throughout the present work, we will make use of the following numerical values for each mass: $(m_u,m_d,m_s)=(5,5,100)$ MeV~\cite{Nakamura:2010zzi}, taking into account isospin symmetry and explicit SU(3) flavor-symmetry breaking. $M_f$ is assigned for the nonlocal (momentum-dependent) effective quark mass. The nonlinear expression for the PS-meson fields, $U_5$ reads:
\begin{equation}
\label{eq:U5}
U_5=\exp\left(\frac{i\gamma_5\lambda\cdot\phi}{\sqrt{2}F_\phi} \right),\,\,\,\,
\lambda\cdot\phi=\left(
\begin{array}{ccc}
\frac{1}{\sqrt{2}}\pi^0+\frac{1}{\sqrt{6}}\eta& \pi^+& K^+\\
 \pi^-&-\frac{1}{\sqrt{2}}\pi^0+\frac{1}{\sqrt{6}}\eta& K^0\\
 K^-& \bar{K}^0&-\frac{2}{\sqrt{6}}\eta
\end{array}\right).
\end{equation}
Here, $F_\phi$ and $\lambda$ are the weak-decay constant for the PS meson as a normalization constant and the Gell-Mann matrix. One can easily see from Eq.~(\ref{eq:ECA}) that the nonlinear PS-meson fields and the quarks interact with the nonlocal (derivative) strength $\propto M_f$. For instance, the effective Lagrangian density for the $qq\phi$ vertex can be obtained from EChA as follows:
\begin{equation}
\label{eq:EFFL1}
\mathcal{L}^\mathrm{nonlocal}_{qq\phi}\sim \frac{i}{F_\phi}\bar{q}
\left[\sqrt{M_f(\partial)}\gamma_5(\lambda\cdot\phi)\sqrt{M_f(\partial)} \right]q.
\end{equation}
If we turn off the momentum dependence in $M_f$, it becomes a positive constant, and one can easily obtain the usual local PS-type effective Lagrangian density, 
\begin{equation}
\label{eq:}
\mathcal{L}^\mathrm{local}_{qq\phi}\sim \frac{i\mathcal{C}_{qq\phi}}{F_\phi}\bar{q}
\left[\gamma_5(\lambda\cdot\phi) \right]q\sim ig_{qq\phi}\bar{q}
\left[\gamma_5(\lambda\cdot\phi) \right]q,
\end{equation}
where $\mathcal{C}_{qq\phi}$ stands for a massive constant. Since EChA in Eq.~(\ref{eq:ECA}) contains derivatives for the quark kinetic part as well as the quark effective mass, as suggested in Refs.~\cite{Nam:2006sx}, the gauge-invariant EChA can be easily obtained by imposing the minimal substitution in Eq.~(\ref{eq:ECA}), $\partial_\mu\to D_\mu\equiv \partial_\mu-iV_\mu$, in which $V$ indicates a local vector field: 
\begin{equation}
\label{eq:ECHA2}
\mathcal{S}_\mathrm{eff}[\phi,m_f,V_\mu;\mu]=
-i\mathrm{Sp}\ln\left[i\rlap{\,/}{D}
-\hat{m}_f-\sqrt{M_f(i\rlap{\,/}{D})}
U_5\sqrt{M_f(i\rlap{\,/}{D})} \right],
\end{equation}
In this way, one ensures the gauge invariance of relevant physical quantities, extracted from Eq.~(\ref{eq:ECHA2}). We also note that this gauging procedure is similar to that given in Ref.~\cite{Dorokhov:2000gu}, in which the single-instanton model was employed. Here is one theoretical caveat: When our writing the gauge-invariant EChA as in Eq.~(\ref{eq:ECHA2}), we assumed the gauge connection in a straight-line path, which is most convenient and practical~\cite{Goeke:2007bj}. 

Considering the effective local vertex for $m=0$ in Eq.~(\ref{eq:MO}), the matrix element in the right-hand-side of Eq.~(\ref{eq:PDFDEFI}) for PDF can be evaluated by the three-point functional derivative with respect to the $\phi$ and $V$ in Eq.~(\ref{eq:ECHA2}) with the delta function,
\begin{equation}
\label{eq:DER}
\frac{\delta^3\mathcal{S}_\mathrm{eff}[\phi,m_f,V_\mu;\mu]}{\delta\phi^\alpha(x)\delta\phi^\beta(y)\delta V_\mu(0)}\Big|_{\phi^{\alpha,\beta},V=0},
\end{equation}
where the superscripts $(\alpha,\beta)$ denote the isospin indices for the PS mesons. Simultaneously, we expand the nonlinear PS-meson field $U_5$ in EChA up to $\mathcal{O}(\phi^2)$, since we are interested in the two PS meson fields for the initial and final states as in Eq.~(\ref{eq:PDFDEFI}). After performing these procedures, the analytical result for PDF via NLChQM reads:
\begin{eqnarray}
\label{eq:PDF}
f_\phi(x)
&=&-\frac{iN_c}{2F^2_\phi}\int\frac{d^4k}{(2\pi)^4}\delta(k_a\cdot n-x\,p\cdot n)
\mathrm{Tr}_\gamma\Big[\sqrt{M_b}\gamma_5\sqrt{M_a}S_a
\rlap{/}{n} S_a\sqrt{M_a}\gamma_5\sqrt{M_b}S_b
\cr
&+&(\sqrt{M_{b}}\cdot n)\gamma_5\sqrt{M_a}S_f\sqrt{M_a}\gamma_5\sqrt{M_b}S_b
-\sqrt{M_b}\gamma_5(\sqrt{M_{a}}\cdot n)S_a\sqrt{M_a}\gamma_5\sqrt{M_b}S_b\Big].
\end{eqnarray}
The subscripts $(a,b)$ indicate two different flavors inside the PS meson with the momenta $(k_a,k_b)\equiv(k,k-p)$ as defined in Figure~\ref{FIG1}. We note that the delta function, $\delta(k_a\cdot n-x\,p\cdot n)$ has been convoluted in the integral over $k$ as mentioned above. It is also emphasized that the second and third terms in the square bracket in the right-hand-side of Eq.~(\ref{eq:PDF}) exist only when we consider the momentum-dependent effective quark mass, since those terms are obtained from the functional derivative of the gauge-invariant EChA in Eq.~(\ref{eq:ECHA2}) with respect to $V_\mu$ which is a special feature of the present nonlocal interaction model. In other words, these {\it derivative} or {\it nonlocal} contributions ensure the gauge invariance for PDF by construction~\cite{Nam:2006sx}. The quark propagator for the flavor $a$ is denoted by
\begin{equation}
\label{eq:PRO}
S_a\equiv\frac{[\rlap{/}{k}_a+(m_a+M_f)]}{k^2_a-(m_a+M_a)^2+i\epsilon}
=\frac{[\rlap{/}{k}_a+\bar{M}_f]}{k^2_a-\bar{M}_a^2+i\epsilon}.
\end{equation}
In the above equation, we introduced a notation $\bar{M}_a$ and indicated the Feynman $\epsilon$ explicitly. The relevant mass functions in Eq.~(\ref{eq:PDF}) read 
\begin{equation}
\label{eq:MDEQM}
M_a=M_0\left[\frac{\mu^2}{k^2_a-\mu^2+i\epsilon} \right]^2,\,\,\,\,
\sqrt{M_a}_\mu=-\sqrt{M_a}\frac{2k_{a\mu}}{(k^2_a-\mu^2+i\epsilon)}.
\end{equation}
Note that we employ the Lorentzian-type structure function for the effective quark mass, as motivated by the instanton physics~\cite{Diakonov:2002fq} as well as employed in Refs.~\cite{Praszalowicz:2001wy,Nam:2006sx,Nam:2006au}. Here, $M_0$ indicates the constituent-quark mass at zero virtuality which is the model parameter to be determined phenomenologically in the next Section.  Evaluating the trace over each spin index and employing the light-cone coordinate vector manipulations~\cite{Praszalowicz:2001wy,Nam:2006sx,Nam:2006au}, 
\begin{equation}
\label{eq:MA}
k\cdot n=k^+=xp^+,\,\,\,\,k^2=k^+k^--k^2_T,\,\,\,\,p^2=m^2_\phi,\,\,\,\,
k\cdot p=\frac{p^+k^-+k^+p^-}{2},\,\,\,\,p^-=\frac{m^2_\phi}{p^+},
\end{equation}
we are left with the contour and polar integrals for $k^-$ and $k_T$, after the delta function integral over $k^+$ in Eq.~(\ref{eq:PDF}):
\begin{eqnarray}
\label{eq:PDF2}
f_\phi(x)&=&-\frac{iN_c}{4F^2_\phi}\int\frac{dk^-d^2k_T}{(2\pi)^3}\left[\mathcal{F}_\mathrm{L}(k^-,k^2_T)+\mathcal{F}_\mathrm{NL,a}(k^-,k^2_T)+\mathcal{F}_\mathrm{NL,b}(k^-,k^2_T) \right]+(x\leftrightarrow\bar{x}),
\end{eqnarray}
where the relevant functions $\mathcal{F}_{\mathrm{L,NL,(a,b)}}$ in the square bracket are explicitly defined in Appendix. We have also used a notation $\bar{x}\equiv(1-x)$ here. The integral in Eq.~(\ref{eq:PDF2}), however, is not simple in comparison to those of usual local models. Since $M_{a,b}$ in Eq.~(\ref{eq:MDEQM}), which can have poles in the denominator during the integrals over $k^-$, appear in the denominator as well as the numerator in the integrand of Eq.~(\ref{eq:PDF2}) simultaneously. For instance, we have to solve the septic equation of $k^-$ in the present work to find the poles in performing a contour integral over $k^-$ with an appropriate cut. The choice of the cut is discussed in detail in Ref.~\cite{Praszalowicz:2001wy}, and we follow their cut scheme.  It is worth mentioning that the numbers of the poles appearing in the calculation of PDF relates to the power of $M_f$ in Eq.~(\ref{eq:MDEQM}). For instance, if we take $M_f=M_0[\cdots]^\mathrm{1\,or\,3}$, it is necessary to solve a quintic or nonic equation to find the poles in the present nonlocal model. As shown in Ref.~\cite{Nam:2006sx}, the change of the power can make effects on the shape of the curves. We will, however, only consider the case with Eq.~(\ref{eq:MDEQM}) in the present work for simplicity, since the changes of the shapes due to the different power appear only around the end points, $x=0$ and $x=1$,~\cite{Nam:2006sx} and the changes can be absorbed in the DGLAP evolution qualitatively. After the contour integral over $k^-$, the polar integrals over $k_T$ can be done easily numerically, and one is led to the final result for PDF (or VQDF).
\section{Numerical results and discussions}
\begin{figure}[t]
\begin{tabular}{cc}
\includegraphics[width=8.5cm]{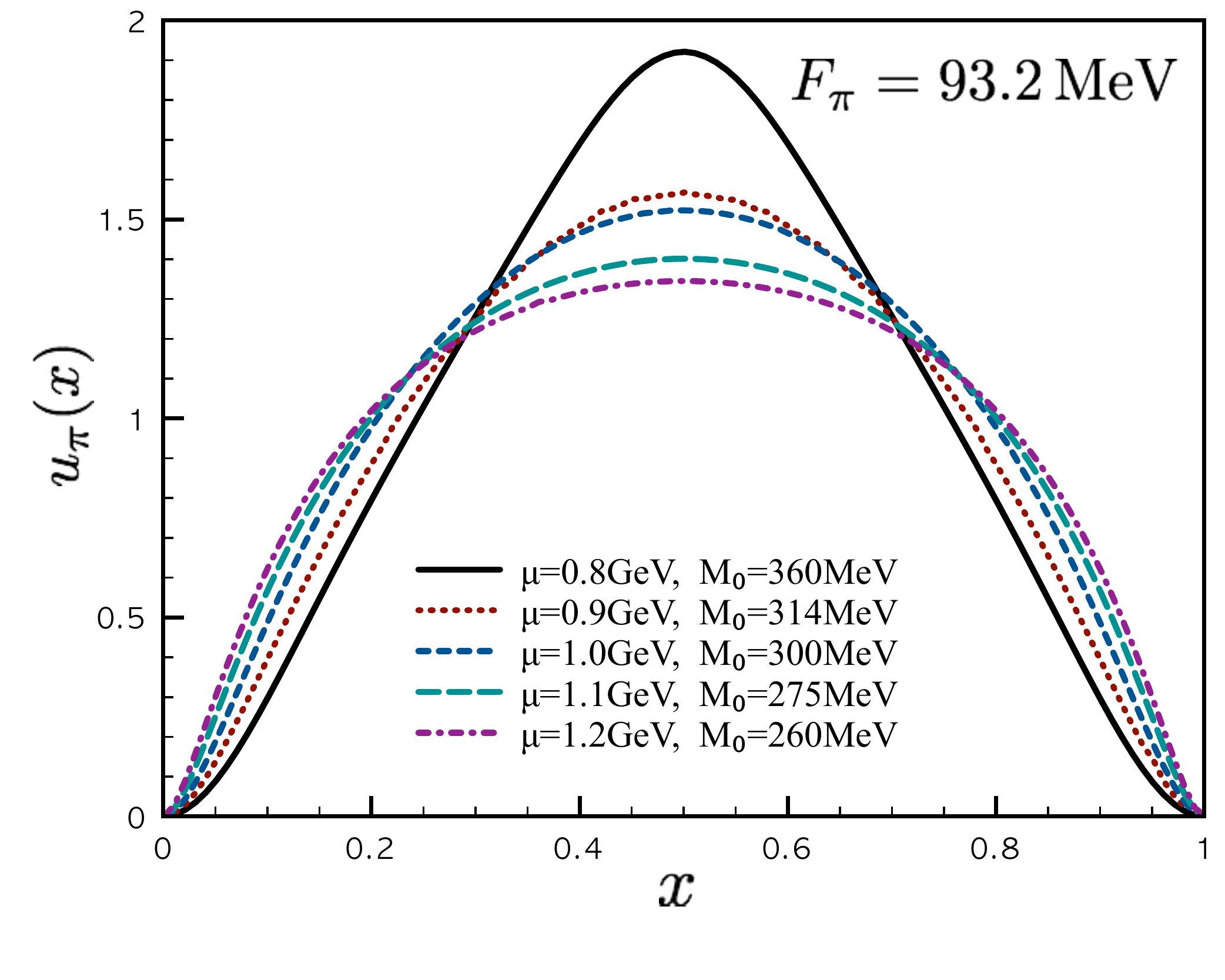}
\includegraphics[width=8.5cm]{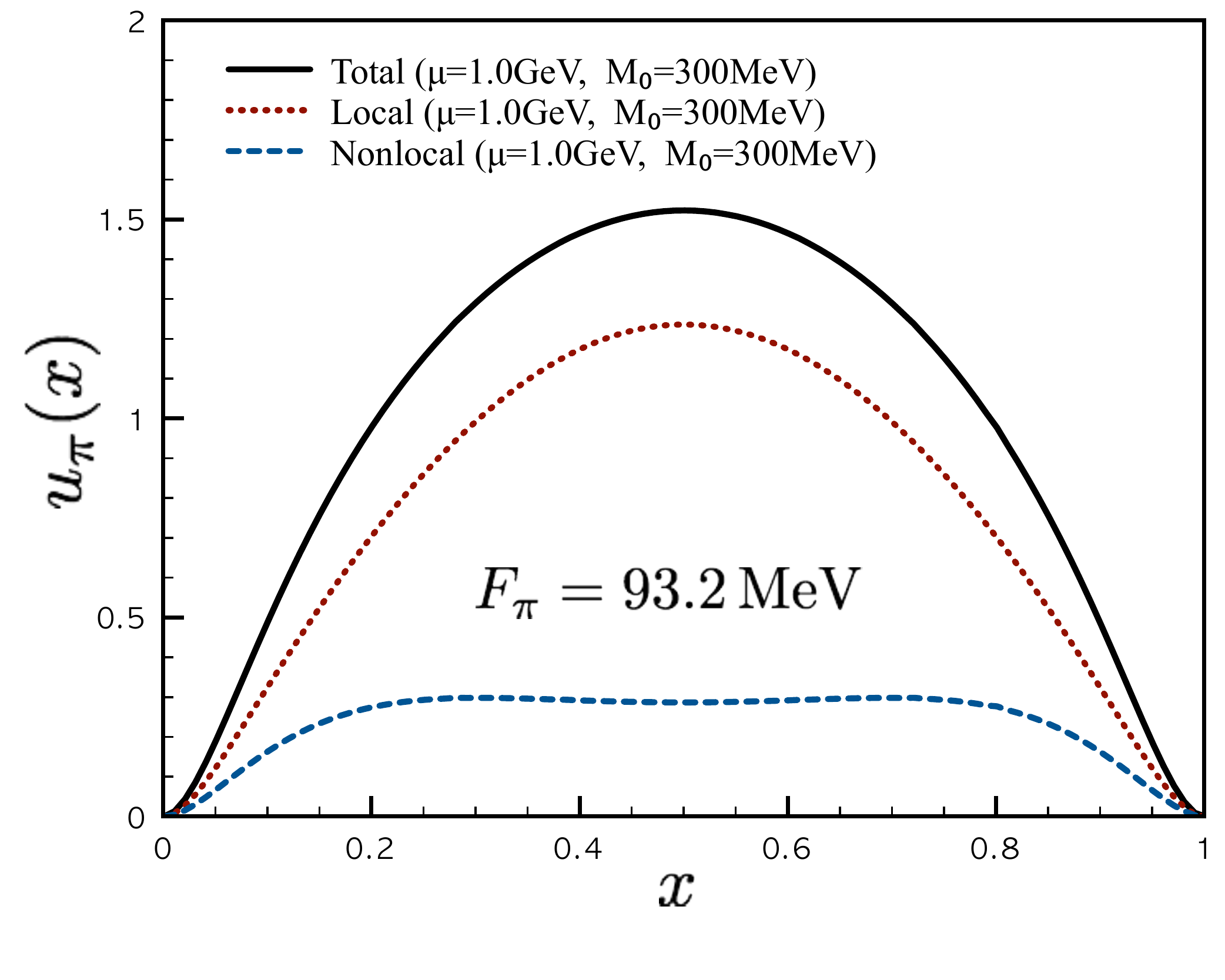}
\end{tabular}
\caption{(Color online) Left: PDF for the pion, $u_\pi(x)$ for various renormalization scale $\mu$ and constituent-quark mass at zero virtuality $M_0$, satisfying the PDF normalization condition Eq.~(\ref{eq:MO1}) with $F_\pi=93.2$ MeV. Right: $u_\pi(x)$ is shown separately for the total (solid), local (dot), and nonlocal (dash) contributions for $(\mu,M_0)=(1,0.3)$ GeV.}       
\label{FIG2}
\end{figure}

In this Section, we provide the numerical results with relevant discussions. First, we would like to explain how to determine the model parameters, saying, the model renormalization scale $\mu$ and the constituent-quark mass at zero virtuality $M_0$ in Eq.~(\ref{eq:MDEQM}). For this purpose, we make use of the PDF normalization condition in Eq.~(\ref{eq:MO1}) and the empirical values for the PS-meson weak-decay constants, $F_{\pi,K}=(93.2,113.4)$ MeV. Note that $F_\phi$ appears in Eq.~(\ref{eq:PDF2}) as a normalization constant in the denominator. Considering the phenomenological scale for the hadrons around $1$ GeV, first, we try $\mu=(0.8\sim1.2)$ GeV to satisfy the normalization condition with $F_\pi=92.3$ MeV as an input for the pion PDF, by varying the $M_0$ value. Corresponding  numerical results for $u_\pi(x)$ are shown in the left panel of Fig.~\ref{FIG2}. As for the various renormalization scales, $M_0$ varies from $260$ MeV to $360$ MeV to satisfy the normalization condition. Moreover, depending on the parameter set of $(\mu,M_0)$, the shape of PDF also changes significantly. We observe a tendency that as $\mu$ increases (and vice versa for $M_0$), PDF gets broadened. In order to choose one parameter set out of the five as given in the left panel of Fig.~\ref{FIG2}, we also consider the kaon case similarly. We compute $F_K\equiv F^\mathrm{computed}_K$, using those parameter sets and Eq.~(\ref{eq:MO1}), then compare them with its empirical value. One can choose the best parameter set, which minimize the quantity, defined as 
\begin{equation}
\label{eq:DEL}
\Delta F_K=|F^\mathrm{computed}_K-F^\mathrm{empirical}_K|.
\end{equation}

We list the numerical results for $F_K$ in Table~\ref{TABLE1}. As shown in the table, the parameter set $(\mu,M_0)=(1,\,0.3)$ GeV presents the smallest deviation $\sim7\%$ from the empirical value of $F_K$. Although we note that the deviation can be reduced by tuning the current-quark masses $(m_{u,d},m_s)$, being different from $(5,100)$ MeV, we do not perform those tunings, since they only produce qualitatively negligible changes in the numerical results. Thus, we will choose this parameter set for all the numerical calculations hereafter. 
\begin{table}[b]
\begin{tabular}{c|ccccc}
$\mu$ [GeV]&$0.8$&$0.9$&$\bf{1.0}$&$1.1$&$1.2$\\
$M_0$ [MeV]&$360$&$314$&$\bf{300}$&$275$&$260$\\
\hline
$F^\mathrm{computed}_K$ [MeV]&$126.1$&$122.3$&$\bf{121.7}$
&$122.2$&$123.0$\\
$\Delta F_K$ [MeV]&$12.7$&$9.0$&$\bf{8.3}$&$8.8$&$9.6$\\
Deviation [$\%$] &$11.2$&$7.9$&$\bf{7.3}$&$7.8$&$8.5$\\
\end{tabular}
\caption{Computed $F_K$ with various parameter sets ($\mu$,$M_0$), which are determined by Eq.~(\ref{eq:MO1}) and the input $F_\pi=93.2$ MeV. We also list the values for $\Delta F_K$ in Eq.~(\ref{eq:DEL}) and deviations from its empirical value. We will use the third parameter set, written in a boldface, for all the numerical results.}
\label{TABLE1}
\end{table}
Using the parameter set chosen, we show the numerical results for $u_\pi(x)$ for the total (solid), local (dot), and nonlocal (dash) contributions separately in the right panel of Fig.~\ref{FIG2}. We observe that the local contribution has a peak $x=0.5$, whereas the nonlocal one show bumps around $x=0.3$ and $0.7$. By integrating the local and nonlocal contributions over $x$, we have about $0.76$ and $0.24$, respectively. The ratio $0.24/0.76\sim1/3$ is a typical value for each contributions in the NLChQM calculations, when the gauge invariance is taken into account explicitly. 

\begin{figure}[t]
\begin{tabular}{cc}
\includegraphics[width=8.5cm]{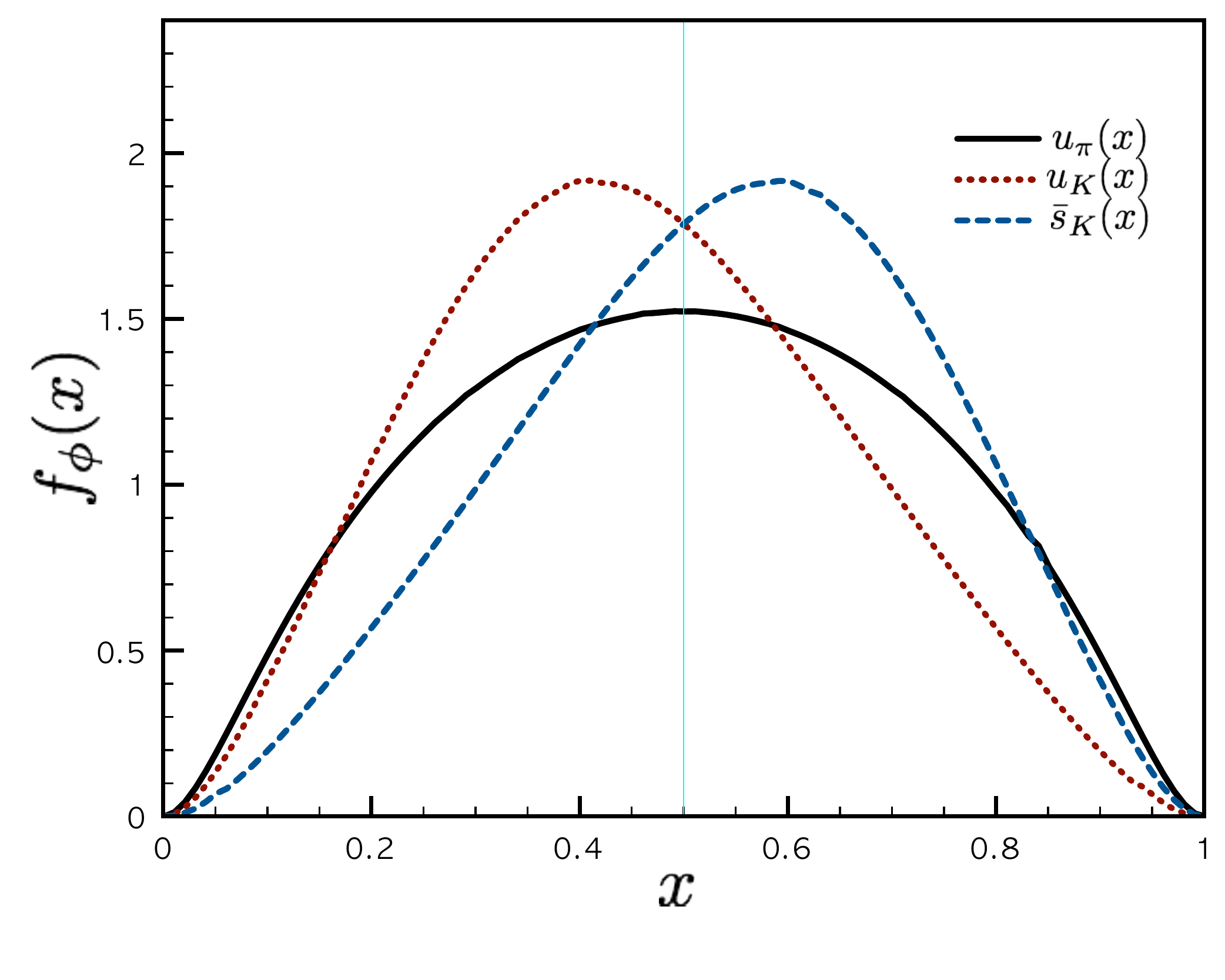}
\includegraphics[width=8.5cm]{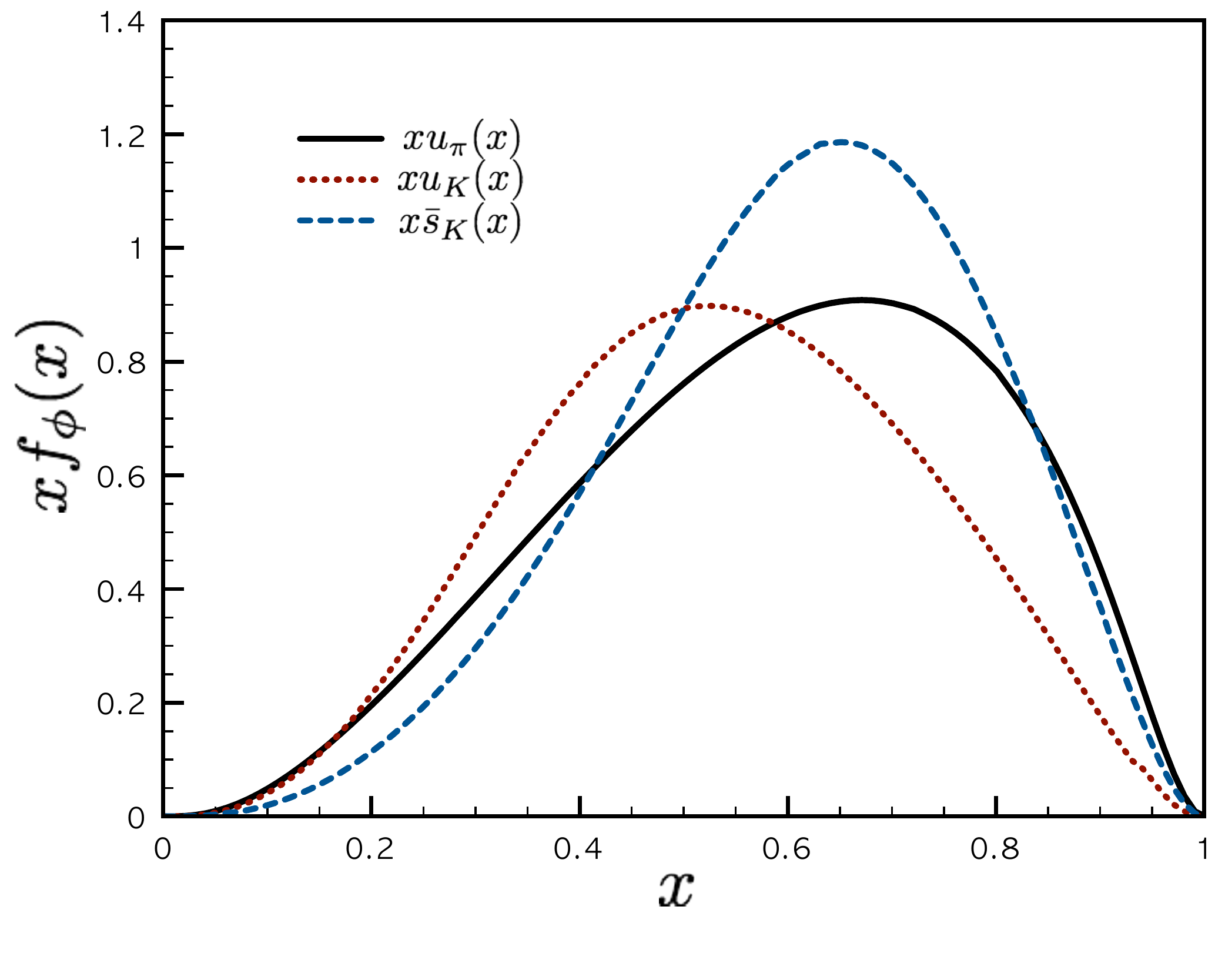}
\end{tabular}
\caption{(Color online) Left: Valance-quark parton distribution functions (VQDF), $u_\pi(x)$ (soild),  $u_K(x)$ (dot), and $\bar{s}_K(x)$ (dash) from Eq.~(\ref{eq:PDF2}). The vertical lines indicate the position for the momentum fraction $x=0.5$. Here, we use $(\mu,M_0,F_\pi,F_K)\approx(1000,300,93.2,121.7)$ MeV and in order to satisfy the normalization condition for PDFs in Eq.~(\ref{eq:MO1}). Right: Those multiplied by $x$, i.e. $xu_\pi(x)$, $xu_K(x)$, and $x\bar{s}_K(x)$ are also given in the same manner.}       
\label{FIG3}
\end{figure}

In the left panel of Fig.~\ref{FIG3}, we draw the numerical results for $u_\pi(x)$ (solid), $u_K(x)$ (dot), and $\bar{s}_K(x)$ (dash). Due to the far heavier mass of the strange-quark mass $\sim100$ MeV, the curves for the kaon present obviously asymmetric shapes, signaling the explicit flavor-SU(3)-symmetry breaking, while that for the pion is clearly symmetric with respect to $x$ according to the isospin symmetry. Since, as in many literatures to investigate PDFs, it is convenient to use the form $xf_\phi(x)$ rather than $f_\phi(x)$~\cite{Ito:2009zc,Matevosyan:2010hh,Nam:2011hg,Nam:2012af}, and the numerical results for those functions are given in the right panel of Fig.~\ref{FIG3}. We note that all the results are qualitatively compatible with other theoretical results~\cite{Shigetani:1993dx,Dorokhov:2000gu,Ito:2009zc,Nguyen:2011jy}. Moreover, the shapes of the curves are different from the previous works~\cite{Nam:2011hg,Nam:2012af}, indicating that the explicit considerations on the momentum-dependent quark masses in the calculations are crucial, in addition to the gauge invariance. 

\begin{figure}[t]
\begin{tabular}{cc}
\includegraphics[width=8.5cm]{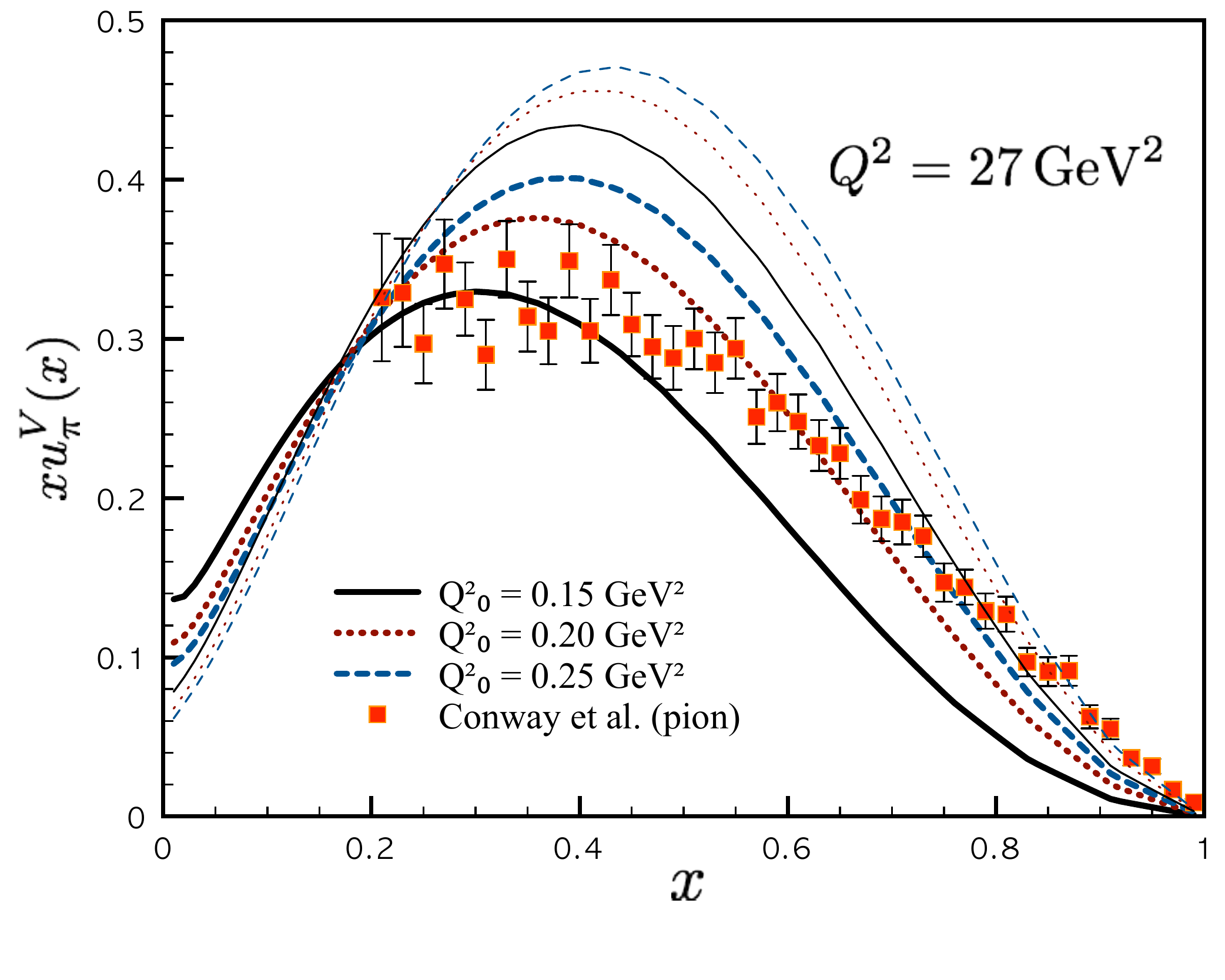}
\includegraphics[width=8.5cm]{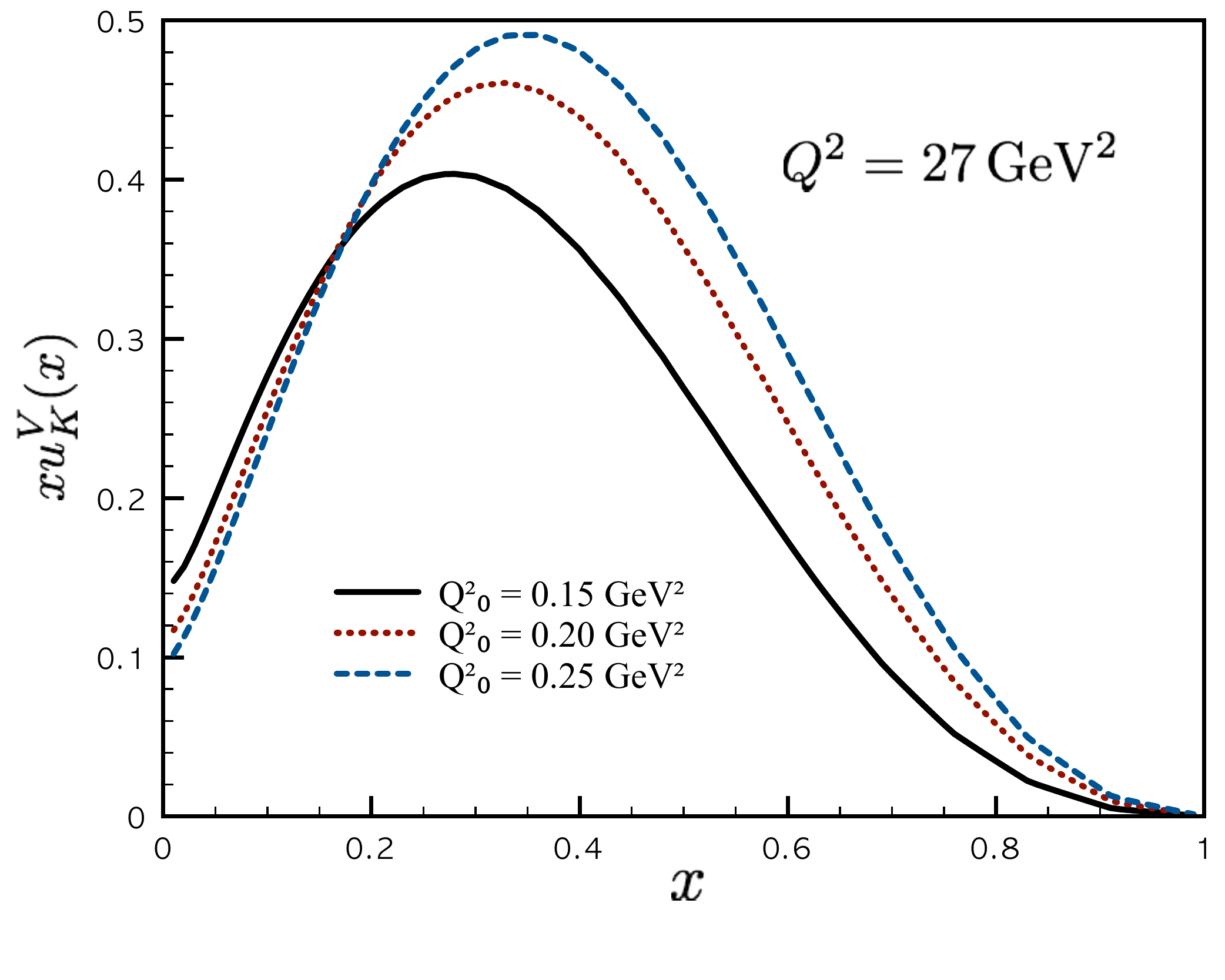}
\end{tabular}
\caption{(Color online) Left: Valance $u$-quark distribution function multiplied by $x$, $xu^V_\pi(x)$ evolved to $Q^2=27\,\mathrm{GeV}^2$ for various initial values $Q^2_0=(0.15,\,0.20,\,0.25)\,\mathrm{GeV}^2$ in (solid, dot, dash) lines, using the LO (thin) and NLO (thick) DGLAP evolutions. Empirical data are taken from the muon-pair production experiment by $252$ GeV pions on tungsten target~\cite{Conway:1989fs}. Right: The same for $xu^V_K(x)$ with the NLO DGLAP evolution.}        
\label{FIG4}
\end{figure}

The parameterization of PDF in the following analytic form is very useful for various applications in data analyses and high-$Q^2$ evolutions:
\begin{equation}
\label{eq:PARAPDF}
xf_\phi(x)=a_\phi\,x^{b_\phi}(1-x)^{c_\phi},
\end{equation}
where the fitting parameters $(a,b,c)_\phi$ are certain positive-real values. Using the numerical results shown in the right panel of Fig.~\ref{FIG3}, those parameters can be fitted and are listed in Table.~\ref{TABLE2} for the low-renormalization scale $\mu=1$ GeV. 
\begin{table}[b]
\begin{tabular}{c|ccc}
&$\hspace{0.3cm}a_\phi\hspace{0.3cm}$
&$\hspace{0.3cm}b_\phi\hspace{0.3cm}$
&$\hspace{0.3cm}c_\phi\hspace{0.3cm}$\\
\hline
$u_\pi$&$7.60$&$2.16$&$1.13$\\
$u_K$&$18.90$&$2.43$&$2.00$\\
$\bar{s}_K$&$35.67$&$3.47$&$1.83$\\
\end{tabular}
\caption{Parameters $(a,b,c)_\phi$ for the pion and kaon PDFs multiplied by $x$ in Eq.~(\ref{eq:PARAPDF}).}
\label{TABLE2}
\end{table}
Using these parameterized PDFs in Eq.~(\ref{eq:PARAPDF}) and Table~\ref{TABLE2} as inputs, we perform the high-$Q^2$ evolution, using the DGLAP equation, to compare our results with the empirical data and theoretical estimations in Refs.~\cite{Conway:1989fs,Aicher:2010cb,Martinelli:1987bh,Sutton:1991ay}. The fortran code {\it QCDNUM} is used to this end~\cite{Botje:2010ay,DGLAP}. In the left panel of Fig.~\ref{FIG4}, we show $xu^V_\pi(x)$, evolved to $Q^2=27\,\mathrm{GeV}^2$ using the LO (thin) and NLO (thick) DGLAP evolutions. Here, the empirical data for the pion are taken from the muon-pair production experiment by $252$ GeV pions on tungsten target~\cite{Conway:1989fs}. Since it is rather uncertain to choose the initial momentum scale $Q_0$ in the DGLAP evolution, we try three different values for it as $Q^2_0=(0.15,0.20,0.25)\,\mathrm{GeV}^2$. Note that these values are related to the momentum scale about $(400\sim500)$ MeV, which are also compatible to typical nonperturbative scales $\Lambda=(0.5\sim0.6)$ GeV in the instanton model~\cite{Shuryak:1982dp,Diakonov:2002fq} and $\Lambda\approx0.4$ GeV in the generic NJL model~\cite{Bowler:1994ir}. Again, these values are very close to those used in Refs.~\cite{Gluck:1991ey,Shigetani:1993dx,Nguyen:2011jy} for the same purpose. As shown in the left panel of Fig.~\ref{FIG4}, the numerical results reproduce the empirical data qualitatively well for the NLO DGLAP evolution, whereas the LO DGLAP evolution results in general show larger than the empirical data. Since the NLO DGALP results are much more compatible to the data, we will present the numerical results only for the NLO ones hereafter. Similarly, we present the numerical results for $xu^V_K(x)$, evolved to $Q^2=27\,\mathrm{GeV}^2$, in the right panel of Fig.~\ref{FIG4}, although there have been no empirical data to be compared to. In comparison to $xu^V_\pi(x)$, the peak positions are shifted to the lower $x$ region, due to the explicit flavor-SU(3)-symmetry breaking, as understood in the right panel of Fig.~\ref{FIG3}. Although there are strength differences in the VQDF curves for each $Q^2_0$ values, the curve shapes are qualitatively similar to each other.   

\begin{figure}[t]
\begin{tabular}{cc}
\includegraphics[width=8.5cm]{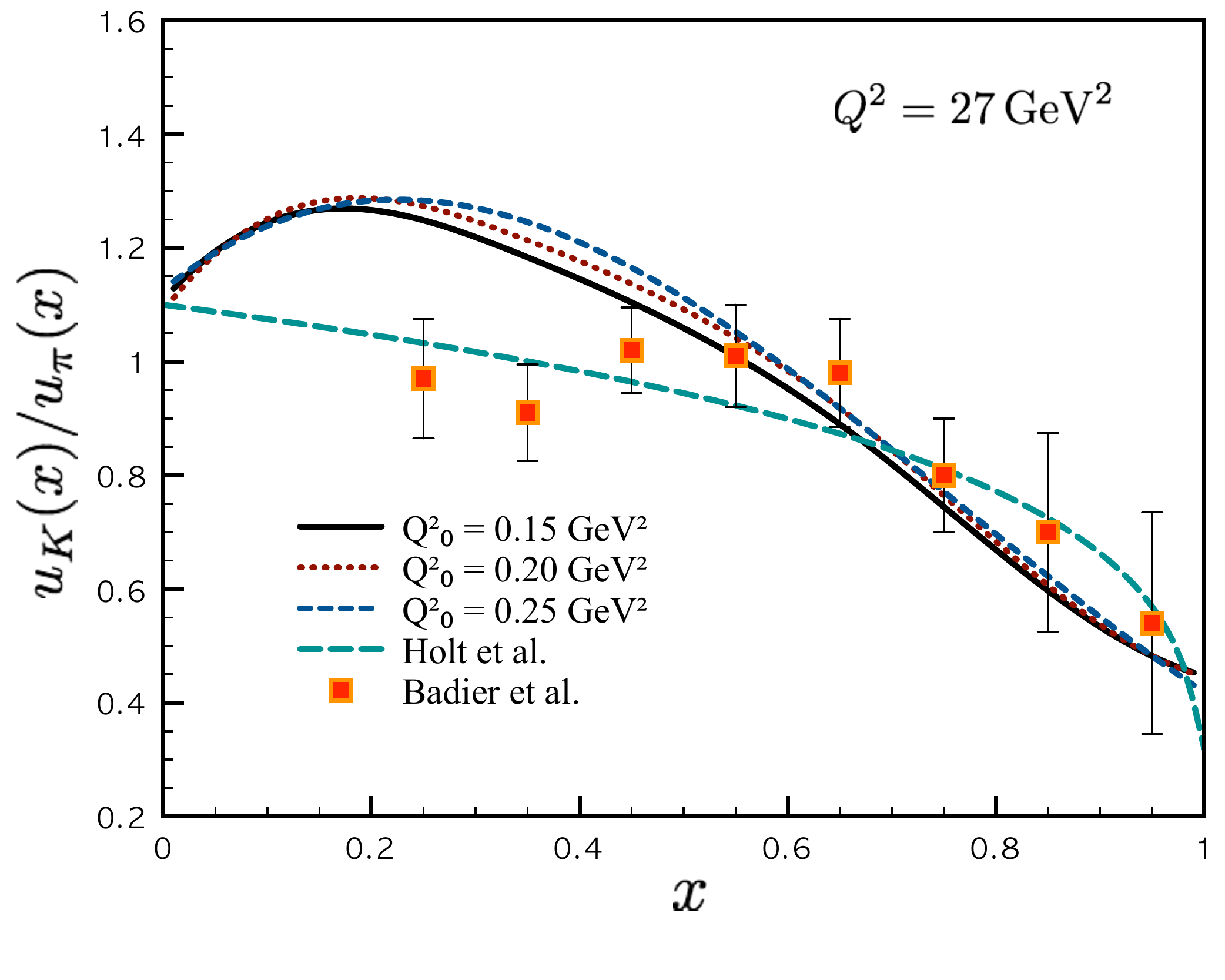}
\includegraphics[width=8.5cm]{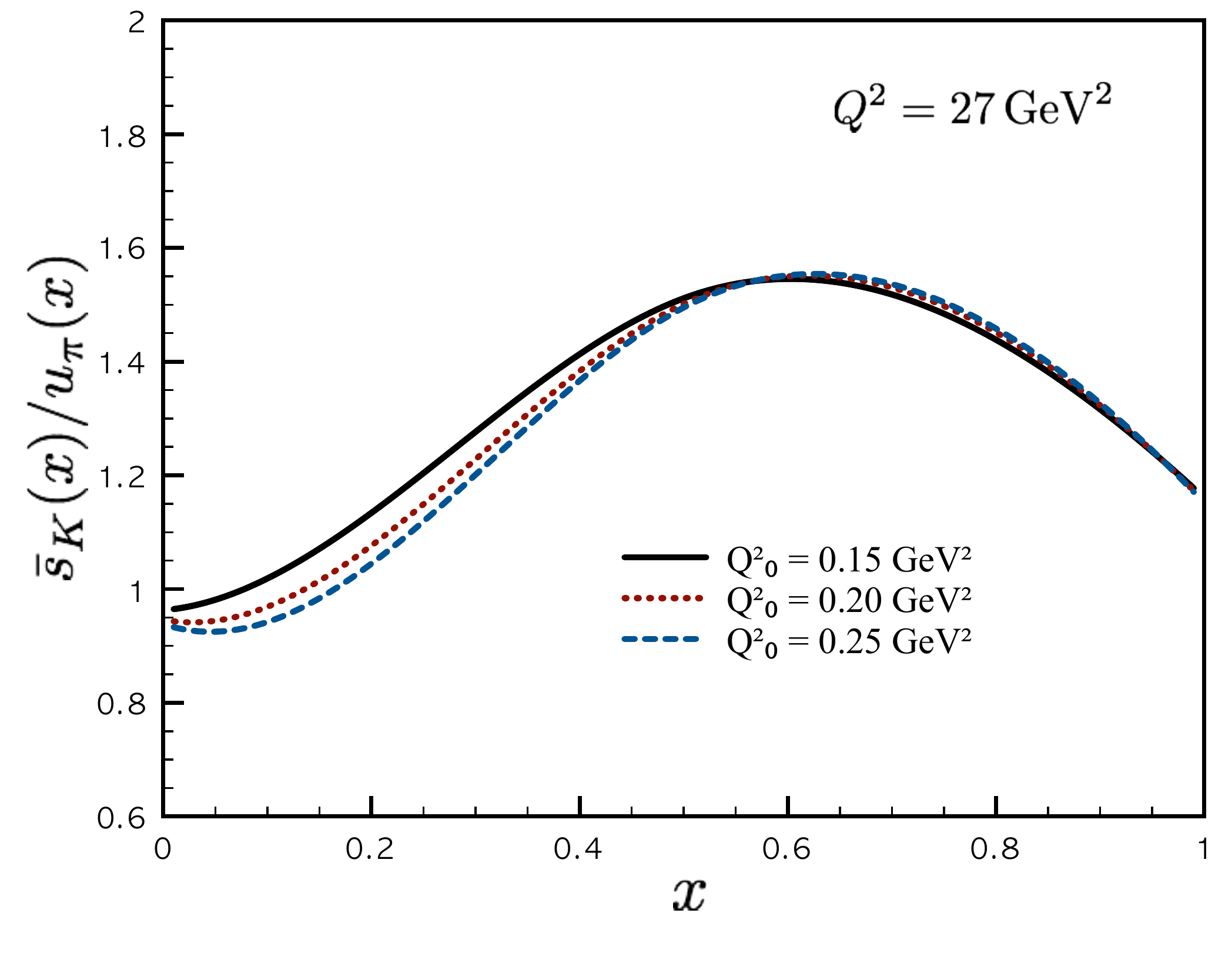}
\end{tabular}
\caption{(Color online) Right: Ratio $u_K/u_\pi$ at $Q^2=27\,\mathrm{GeV}^2$ as a function of $x$ for various initial values $Q^2_0=(0.15,\,0.20,\,0.25)\,\mathrm{GeV}^2$ in (solid, dot, dash) lines. We also show the fitted curve in Ref.~\cite{Holt:2010vj} (long dash). Empirical data taken from the $150\,\mathrm{GeV}$ incident-beam experiment for $(K^-,\pi^-)+\mathrm{nucleus}\to\mu^+\mu^-X$~\cite{Badier:1980jq}. Right: The same for $\bar{s}_K(x)/u_K(x)|$.}       
\label{FIG5}
\end{figure}

Now, we present the numerical results for the ratio of two different PDFs, i.e. $u_\pi(x)/u_K(x)$ at $Q^2=27\,\mathrm{GeV}^2$ for the three different initial values $Q^2_0=(1.5,2.0.2.5)\,\mathrm{GeV}^2$ (solid, dot, dash) in the left panel of Fig.~\ref{FIG5}. They can be compared with the data from the $150\,\mathrm{GeV}$ incident-beam experiment for $(K^-,\pi^-)+\mathrm{nucleus}\to\mu^+\mu^-X$~\cite{Badier:1980jq}. We also depict the fitted curve for the data: $u_K(x)/u_\pi(x)=1.1(1-x)^{0.22}$ (long dash)~\cite{Holt:2010vj}. Although the numerical results match with the data qualitatively well in the region $x\gtrsim0.5$, we observe overshoots for the smaller $x$ by $(20\sim30)\%$. Moreover, there are visible deviations between the numerical results and the fitted curve of Ref.~\cite{Holt:2010vj}. We note that the curve shapes of our results resemble that from the reduced Bethe-Salpeter equation (BSE) vertex calculation~\cite{Nguyen:2011jy}, although the strengths are slightly different. Except for the region $x\lesssim0.2$, the result given in Ref.~\cite{Shigetani:1993dx} is also similar to ours in shape. In the right panel of Fig.~\ref{FIG5}, the numerical results for $\bar{s}_K(x)/u_K(x)$ are given in the same manner with the left one. Interestingly, the curve behavior is quite different from that for $u_K(x)/u_\pi(x)$. Again, this discrepancy can be understood by the explicit flavor-SU(3)-symmetry breaking effects. As observed by comparing the curves in Figs.~\ref{FIG3} and \ref{FIG4}, $f^V_\phi(x)$ in the vicinity of $x=1$ is almost unaffected by the DGLAP evolution. Hence, as discussed in Refs.~\cite{Shigetani:1993dx,Nguyen:2011jy}, $f^V_\phi(x\to1)$ can be considered to represent the relatively pure nonperturbative QCD contributions. In the previous work~\cite{Nam:2012af}, the ratio of $u_K/u_\pi$ was approximated for the leading local contribution as
\begin{equation}
\label{eq:REL}
\frac{u_K(x)}{u_\pi(x)}\Big|_{x\to1}\approx\frac{F^2_\pi}{F^2_K}\left[1-\frac{m_s-m_u}{M_0} \right]\approx0.4.
\end{equation}
where we have used relevant quantities employed and computed in the present work, i.e. 
\begin{equation}
\label{eq:VAL}
(F_\pi,F_K,M_0,m_u,m_s)=(93.2,121.7,300,5,100)\,\mathrm{MeV}.
\end{equation}
Considering that the nonlocal contribution produces $(20\sim30)\%$ additional strength, we have approximately $u_K(x\to1)/u_\pi(x\to1)=(0.52\sim0.53)$, which is comparable with the actual value $0.45$. It is worth mentioning that different forms for the ratio were suggested as $(M_u/M_s)^2$ from the NJL calculation~\cite{Shigetani:1993dx} and $(F_\pi/F_K)(M_u/M_s)^4$ from the BSE calculation~\cite{Nguyen:2011jy}.

\begin{figure}[t]
\begin{tabular}{cc}
\includegraphics[width=8.5cm]{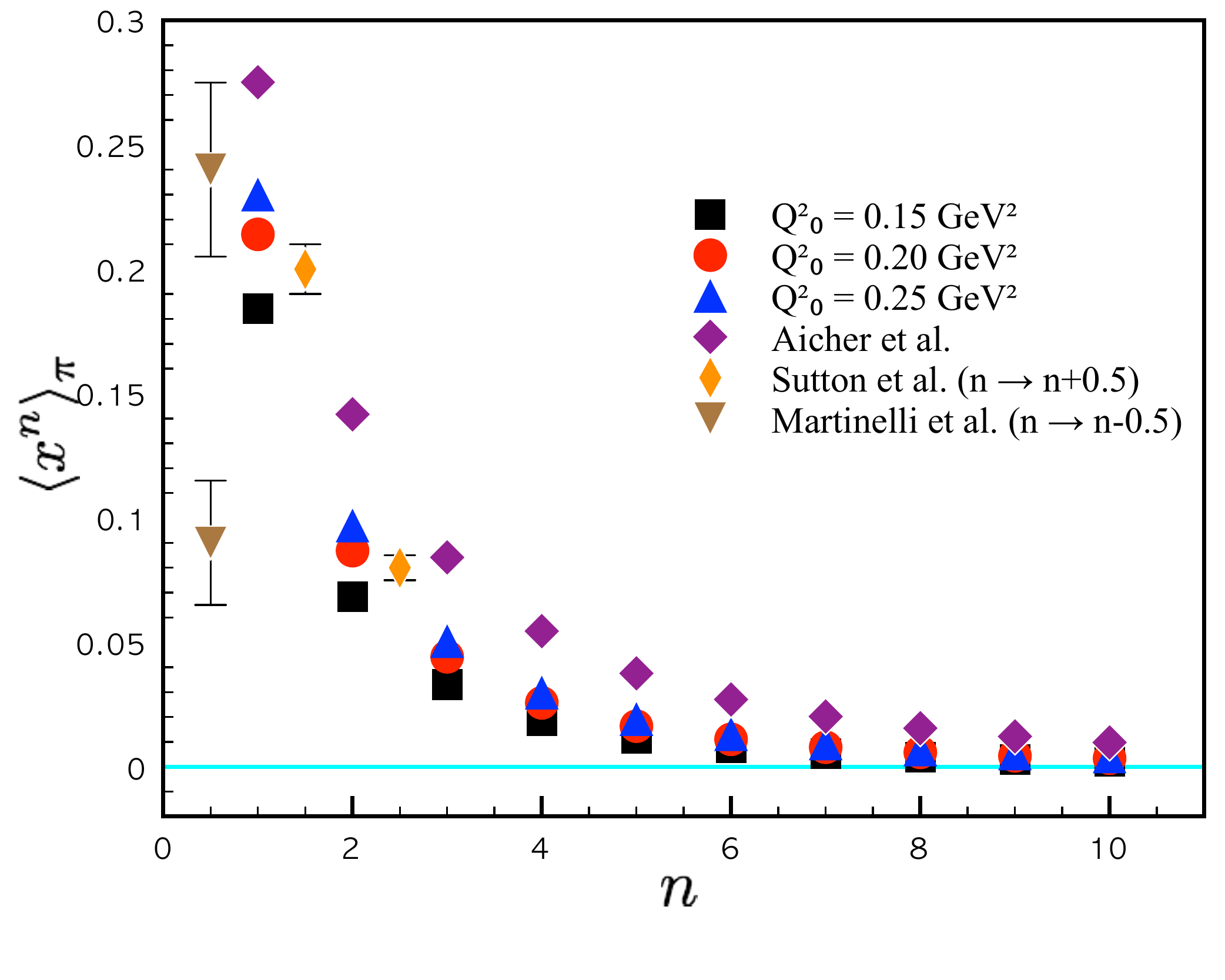}
\includegraphics[width=8.5cm]{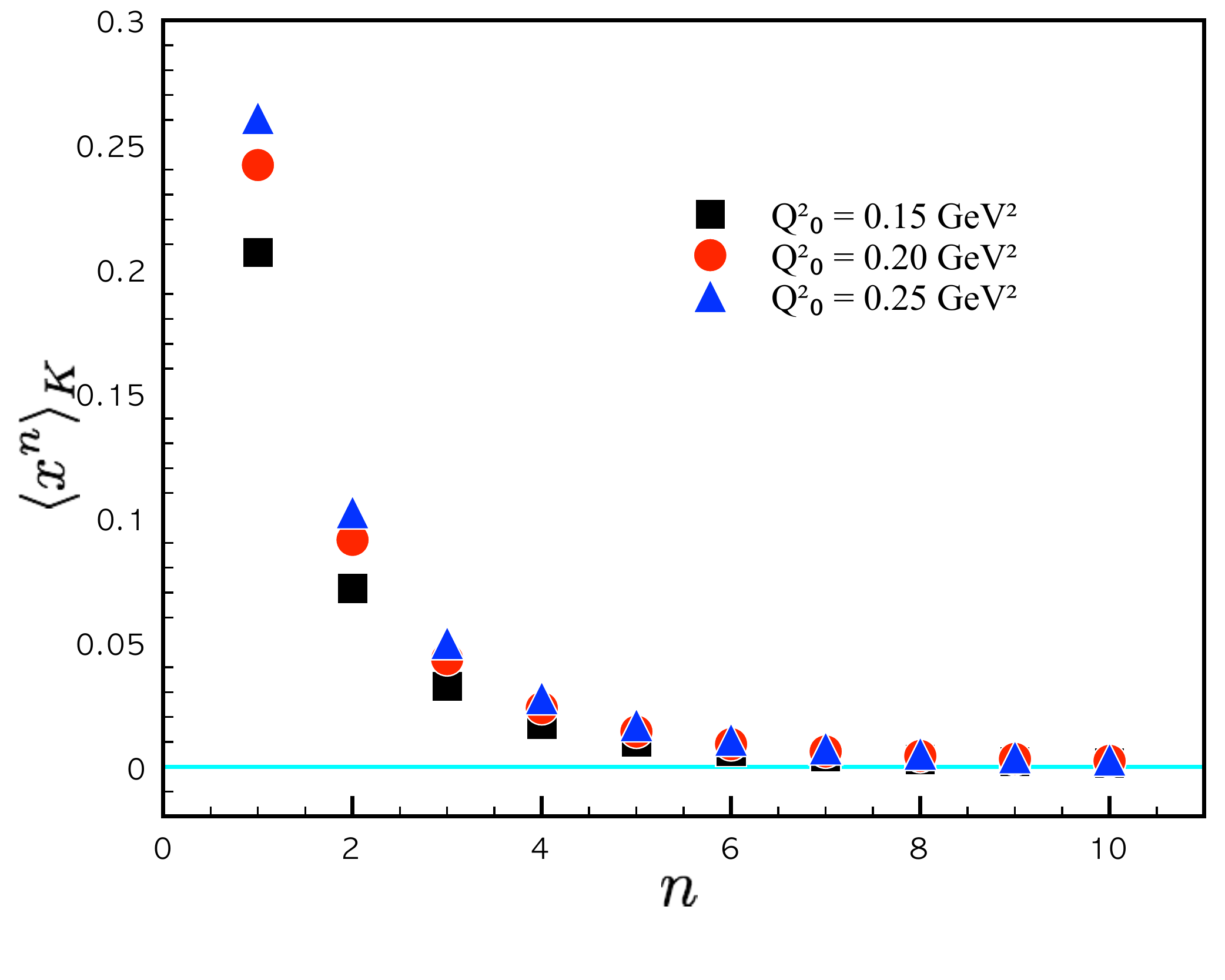}
\end{tabular}
\caption{(Color online) Left: $n$-th moments for the pion VQDFs for $Q^2_0=(0.15\sim0.25)\,\mathrm{GeV}^2$ at $Q^2=27\,\mathrm{GeV}^2$. The data are taken from Aicher {\it et al}.~\cite{Aicher:2010cb}, Martinelli {\it et al}.~\cite{Martinelli:1987bh}, and Sutton {\it et al}.~\cite{Sutton:1991ay}. For clearance, we shift the data of Ref.~\cite{Martinelli:1987bh} and Ref.~\cite{Sutton:1991ay} by $-0.5$ and $+0.5$ for $n$, respectively. Right: The same for the kaon. For the numerics, see Table~\ref{TABLE3}.}       
\label{FIG6}
\end{figure}

Finally, we would like to discuss the moments of the pion and kaon VQDFs as defined in Eq.~(\ref{eq:MO1}). For brevity, we will take into account those only  for the $u$ quark. They are very useful to analyze the data and determine unknown parameters in models and the DGLAP evolutions. In Refs.~\cite{Aicher:2010cb,Martinelli:1987bh,Sutton:1991ay}, the first two moments for the pion, $\langle x^{n=1,2}\rangle_\pi$ are given as follows:
\begin{eqnarray}
\label{eq:MOMENTS}
2\langle x\rangle_\pi&\approx&0.55,\,\,\,\,2\langle x^2\rangle_\pi\approx0.18,
\,\,(\mathrm{fit}1)\,\,:\,\,Q^2=4\,\mathrm{GeV}^2~[31],
\cr
2\langle x\rangle_\pi&=&0.40\pm0.02,\,\,\,\,2\langle x^2\rangle_\pi=0.16\pm0.01,
\,\,:\,\,Q^2=4\,\mathrm{GeV}^2~[30],
\cr
2\langle x\rangle_\pi&=&0.46\pm0.07,\,\,\,\,2\langle x^2\rangle_\pi=0.18\pm0.05,
\,\,:\,\,Q^2=49\,\mathrm{GeV}^2~[32].
\end{eqnarray}
The values in Ref.~\cite{Aicher:2010cb} were obtained by the NLO analysis (fit1) of the data for the Drell-Yan process $\pi^- N\to\mu^+\mu^-X$ of the experiments by E-615~\cite{Conway:1989fs} and NA10~\cite{Bordalo:1987cr} collaborations. Ref.~\cite{Sutton:1991ay} also analyzed the data of NA10 collaboration. On the contrary, the moments in Ref.~\cite{Martinelli:1987bh} were obtained by the lattice QCD simulation with the Wilson fermions in the quenched approximation for the pion structure functions. From our numerical results, we have the followings, using the curves depicted in Fig.~\ref{FIG4}:
\begin{equation}
\label{eq:OURS}
2\langle x\rangle_\pi=(0.37\sim0.46),\,\,\,\,2\langle x^2\rangle_\pi=(0.14\sim0.19)
\,\,:\,\, Q^2=27\,\mathrm{GeV}^2
\end{equation}
for $Q^2_0=(0.15,0.20,0.25)\,\mathrm{GeV}^2$ as shown in the left panel of Fig.~\ref{FIG6} in the (square, circle, triangle), respectively. We note that, although the $Q^2$ values are different from other estimations, these values  in Eq.~(\ref{eq:OURS}) are well compatible to those given in Eq.~(\ref{eq:MOMENTS}). The schematic comparison of these moments from different works are given in the left panel of Fig.~\ref{FIG6}. For clearance, we present the data of Refs. ~\cite{Sutton:1991ay} (diamond) and \cite{Martinelli:1987bh} (nabla), shifted  by $-0.5$ and $+0.5$ for $n$, respectively. Since, in Ref.~\cite{Aicher:2010cb}, the authors provided the parameterized VQDF as $u^V_\pi(x)=0.077\times x^{-0.85}(1-x)^{1.75}(1+89.4x^2)$ at $Q^2=4\,\mathrm{GeV}^2$, we can compute other higher moments beyond the first two as in the left panel of Fig.~\ref{FIG6} (rhombus). As shown there, the present numerical results for the moments are in qualitatively good agreement with other analyses and theories. In the right panel of Fig.~\ref{FIG6}, we also present the moments for the kaon VQDF, as functions of $n$. Overall behavior of the curve is very similar to that for the pion, whereas visible differences are observed for small $n$ values. This tendency can be understood clearly by seeing Table~\ref{TABLE3}, in which we list all the moments for the pion and kaon for those $Q^2_0$ values at $Q^2=27\,\mathrm{GeV}^2$. From the numerics given in the table, we see that the higher moments for $n>2$ are very small and almost identical for the two mesons. As a consequence, we have the following observation:
\begin{equation}
\label{eq:XXX}
\frac{\langle x^{n=1,2}\rangle_K}{\langle x^{n=1,2}\rangle_\pi}\approx1.1,\,\,\,\,
\frac{\langle x^{n>2}\rangle_K}{\langle x^{n>2}\rangle_\pi}\approx1.
\end{equation}
\begin{table}[b]
\begin{tabular}{c|c|cccccccccc}
&$Q^2_0$ [GeV$^2$]&$n=1$&$2$&$3$&$4$&$5$&$6$&$7$&$8$&$9$&$10$\\
\hline
\hline
\multirow{3}{*}{$\langle x^n\rangle_\pi$}&$0.15$
&$0.184$&$0.068$&$0.033$&$0.018$&$0.011$
&$0.007$&$0.005$&$0.004$&$0.003$&$0.002$\\
&$0.20$
&$0.214$&$0.087$&$0.044$&$0.026$&$0.016$
&$0.011$&$0.008$&$0.006$&$0.004$&$0.003$\\
&$0.25$
&$0.230$&$0.097$&$0.050$&$0.030$&$0.019$
&$0.013$&$0.009$&$0.007$&$0.005$&$0.004$\\
\hline
\multirow{3}{*}{$\langle x^n\rangle_K$}&$0.15$
&$0.207$&$0.072$&$0.032$&$0.017$&$0.010$
&$0.006$&$0.004$&$0.003$&$0.002$&$0.002$\\
&$0.20$
&$0.242$&$0.091$&$0.043$&$0.024$&$0.014$
&$0.009$&$0.006$&$0.004$&$0.003$&$0.002$\\
&$0.25$
&$0.261$&$0.102$&$0.050$&$0.028$&$0.017$
&$0.011$&$0.007$&$0.005$&$0.004$&$0.003$\\
\end{tabular}
\label{TABLE3}
\caption{$n$-th moments for the pion and kaon VQDF for $Q^2_0=(0.15\sim0.25)\,\mathrm{GeV}^2$ at $Q^2=27\,\mathrm{GeV}^2$.}
\end{table}

\section{Summary and conclusion}
We have investigated the pion and kaon PDFs employing the OPE technique for the handbag diagram with the twist-2 operator and the gauge-invariant NLChQM, which conserves the vector currents. All the model parameters were determined by the empirical inputs and model-independent constraint, i.e. PS-meson weak-decay constants and the normalization condition for VQDF. Once all the parameters fixed, we performed the high-$Q^2$ evolution by the LO and/or NLO DGLAP equation for the different initial momentum scales. In what follows, we summarize important observations in the present work:
\begin{itemize}
\item In order to satisfy the gauge invariance, we include the {\it nonlocal} contribution to PDF in addition to the leading {\it local} one. The nonlocal contribution provides about $30\%$ strength for PDF, as usual in the similar approaches. Appropriate consideration for the momentum-dependence of the effective quark mass changes the shapes of PDF, in comparison to the simplified calculations as in the previous work. 
\item Taking $F_\pi=93.2$ MeV as an input, we determine the renormalization scale $\mu=1$ GeV as well as the constituent-quark mass at zero virtuality $M_0=300$ MeV, together with the information $F_K=113.4$ MeV. These values give $F_K=121.7$ MeV, which is slightly larger than its empirical value by about $7\%$. Computed PDFs for the pion and kaon at the low-renormalization scale are compatible qualitatively with other theoretical results, such as the single-instanton model, BSE approach, NJL model, and so on. 
\item After performing the DGALP evolution up to $Q^2=27\,\mathrm{GeV}^2$ for three different initial momentum scales $Q^2_0=(0.15,0.20,0.25)\,\mathrm{GeV}^2$, we compare VQDFs for the pion and kaon with the empirical data as well as theoretical calculations, resulting in qualitatively good agreement with them. However, we still observe sizable differences with the empirical data for some specific $x$ regions, especially as shown in $u_K(x)/u_\pi(x)$.
\item It turns out that the moments for the pion and kaon VQDFs are well compatible with experimental and theoretical estimations. We observe that the differences between the higher moments for $n>2$ are almost negligible, whereas the first two moments, i.e. $n=(1,2)$, show a tendency that $\langle x^{n=1,2}\rangle_K/\langle x^{n=1,2}\rangle_\pi\approx1.1$. 
\end{itemize}
Since we are now equipped with PDF in the gauge-invariant manner, it must be interesting to apply this PDF to the investigations for other physical quantities. For instance, we can use the present result to compute the nucleon PDF together with the PS-meson FF, together with an ansatz for the bare nucleon PDF as studied in Refs.~\cite{Edin:1998dz,Alwall:2005xd,Nematollahi:2012zz}. Related works are under progress and appear elsewhere.

\section*{Acknowledgments}
The author thanks C.~W.~Kao and H.~Kohyama for fruitful discussions and comments. He acknowledges the partial support from NCTS (North) of Taiwan.
\section*{Appendix}
The relevant functions in Eq.~(\ref{eq:PDF2}) are defined as follows:
\begin{eqnarray}
\label{eq:FLOCAL}
&&\mathcal{F}_\mathrm{L}=\frac{4p^+\eta^2D^4_b\left[D^4_aD^8_b\left[\hat{x}k^2_T+\bar{x}^2 k^-p^+\right]+2\bar{x}D^4_b\left(D^4_bm_b+\eta \right)\left(D^4_am_a+\eta \right)
+xD^4_a\left(D^4_bm_b+\eta \right)^2\right]}{\left[D^8_a(\xi_a-\alpha k^-)+2m_a\eta D^4_a+\eta^2\right]_a\left[D^8_b[\xi_b-\beta k^-+\delta)]+2m_b\eta D^4_b+\eta^2\right]^2_b},
\cr
&&\mathcal{F}_\mathrm{NL,a}=
\frac{4p^+\eta^2\left[\eta\left(D^4_bm_b+\eta \right)+ D^4_a[m_a\eta
+D^4_b(2k^2_T+\tilde{x}k^-p^++m_am_b+xm^2_\phi)]\right]}{\left[D^8_a(\xi_a-\alpha k^-)+2m_a\eta D^4_a+\eta^2\right]_a\left[D^8_b[\xi_b-\beta k^-+\delta)]+2m_b\eta D^4_b+\eta^2\right]_b}\left(\frac{x}{\gamma-\alpha k^-}\right),
\cr
&&\mathcal{F}_\mathrm{NL,b}=
\frac{4p^+\eta^2\left[\eta\left(D^4_bm_b+\eta \right)+ D^4_a[m_a\eta
+D^4_b(2k^2_T+\tilde{x}k^-p^++m_am_b+xm^2_\phi)]\right]}{\left[D^8_a(\xi_a-\alpha k^-)+2m_a\eta D^4_a+\eta^2\right]_a\left[D^8_b[\xi_b-\beta k^-+\delta)]+2m_b\eta D^4_b+\eta^2\right]_b}\left(\frac{\bar{x}}{\gamma-\beta k^-+\delta} \right),
\end{eqnarray}
where $m_\phi$ represents the PS-meson mass and the various notations read:
\begin{eqnarray}
\label{eq:PARA}
(\bar{x},\,\,\hat{x},\,\,\tilde{x})&=&(1-x,\,\,2-x,\,\,1-2x),
\cr
(\alpha,\,\,\beta,\,\,\gamma,\,\,\delta,\,\,\eta,\,\,\xi_a,\,\,\xi_b)&=&(xp^+,\,\,-\bar{x}p^+,\,\,k^2_T+\mu^2,\,\,-\bar{x}m^2_\phi,\,\,M_0\mu^4,\,\,k^2_T+m^2_a,\,\,k^2_T+m^2_b),
\cr
(D^2_a,\,\,D^2_b)&=&(\gamma-\alpha k^-,\,\,\gamma-\beta k^-+\delta).
\end{eqnarray}
In deriving above equations, we have assumed that $p^2=m^2_\phi=p^+p^--p^2_T\sim p^+p^-$, considering that the transverse momentum for the meson is much smaller than the longitudinal ones. i.e.  $p_T\sim0$ .

In order to perform a contour integral for Eq.~(\ref{eq:FLOCAL}) over $k^-$, we pick the poles from the denominator. We have $[\cdots]_a=\sum^5_{i=0}a_i(k^-)^i$ in the denominator of Eq.~(\ref{eq:FLOCAL}). The coefficients $a_i$ are written as
\begin{eqnarray}
\label{eq:COEF}
a_5&=&\alpha^5,\,\,\,\,
a_4=-\alpha^4(4\gamma+\xi_a),\,\,\,\,
a_3=\alpha^3(6\gamma^2+4\gamma\xi_a),\,\,\,\,
a_2=-\alpha^2(4\gamma^3+6\gamma^2\xi_a+2m_a\eta),\,\,\,\,
\cr
a_1&=&\alpha^1(\gamma^4+4\gamma^3\xi_a+4m_a\gamma\eta),\,\,\,\,
a_0=-\alpha^0(\gamma^4\xi_a+2m_a\gamma^2\eta+\eta^2).
\end{eqnarray}
The poles can be obtained by solving this quintic equation with respect to $k^-$ numerically.


\end{document}